\documentclass[11pt]{article}
\usepackage{amsmath,amsfonts,amssymb,graphicx,fullpage}
\usepackage[numbers,sort&compress]{natbib}

\begin{document}
\begin{center}
\parbox{140mm}{\begin{flushright}{\small NIKHEF/2007-022\\October 10, 2007}\end{flushright}}

\vspace{30mm}
{\LARGE DBI Inflation using a One-Parameter Family of\\Throat Geometries}

\vspace{15mm}
{\large Florian Gmeiner\footnote{fgmeiner@nikhef.nl} and Chris D. White\footnote{cwhite@nikhef.nl}}

\vspace{5mm}
{\it NIKHEF, Kruislaan 409, 1098 SJ Amsterdam, The Netherlands} \\

\vspace{20mm}
{\bf Abstract}\\[2ex]
\parbox{140mm}{
We demonstrate the possibility of examining cosmological signatures in the DBI inflation setup using the BGMPZ solution, a one--parameter
family of geometries for the warped throat which interpolate between the Maldacena--Nu{\~n}ez and Klebanov--Strassler solutions.
The warp factor is determined numerically and subsequently used to calculate cosmological observables including the scalar and
tensor spectral indices, for a sample point in the parameter space.
As one moves away from the KS solution for the throat the warp factor is qualitatively different, which leads to a significant
change for the observables, but also generically increases the non--Gaussianity of the models.
We argue that the different models can potentially be differentiated by current and future experiments.}
\end{center}

\thispagestyle{empty}
\clearpage

\tableofcontents

\section{Introduction}\label{sec:intro}
There has been much interest in recent years in cosmological applications of string theory.
The availability of precision data relating to the cosmic microwave background (CMB) opens up the possibility of constraining
the phenomenology of string--inspired models, in particular regarding the dynamics of the inflationary epoch.
Essentially, signatures of the extreme conditions governing the behaviour of the early universe are amplified by inflationary
expansion and can be accessible by modern day observations.
The typical energy scales involved far exceed those obtainable in particle collider experiments.
In this paper we consider a particular inflation scenario, the Dirac--Born--Infeld (DBI) model
of~\cite{Silverstein:2003hf,Alishahiha:2004eh}.
In the usual formulation of this setup, a D3 brane rolls down a warped throat towards a $\overline{\text{D3}}$ brane where the
interbrane separation is identified with the inflaton field.
The brane moves relativistically, but its speed is curtailed by the warped geometry so that the potential energy dominates and
inflation can occur.
The observable consequences of this model depend of course upon the choice of solution for the warped throat.
The original DBI proposal used a AdS--like geometry, with an artificially imposed cutoff where the throat joins the bulk geometry,
which in principle is unknown although irrelevant for cosmological implications.
Such a geometry is unstable, and more properly one should consider a solution of the supergravity field equations with non--trivial
D--brane fluxes to stabilise the compact geometry by dynamically generating a cut--off.
Not many such solutions are known analytically, but two examples are the Klebanov--Strassler~\cite{Klebanov:2000hb} (KS) and
Maldacena--Nu{\~n}ez~\cite{Maldacena:2000yy} (MN) results.\\

Several investigations of DBI inflation from the KS geometry abound in the
literature~\cite{Kecskemeti:2006cg,Shiu:2006kj,Easson:2007dh,Shandera:2006ax,Bean:2007hc}, including systematic comparisons to
current cosmological data.
In this paper we show that it is also possible to examine the inflationary consequences of the BGMPZ geometries introduced
in~\cite{Butti:2004pk}.
These are a one--parameter family of geometries describing a deformation of the KS solution\footnote{The BGMPZ solutions originally
arose in a different context, that of the AdS / CFT correspondence, as the baryonic branch of the KS solution.}, and
interpolating smoothly between the KS and MN backgrounds.
Although the metric is not known fully analytically, one can solve for it numerically and thus obtain results for numerous observables.
The hope is that these can be compared with experimental data for quantities such as spectral indices, which may then allow one to
extract geometrical information about the throat--geometry.
To this aim, we show examples of how these quantities vary as a function of the BGMPZ parameter $\xi$ which characterises the family
of solutions, at typical points in parameter space. We note that inflationary consequences of these geometries were also considered
in~\cite{Dymarsky:2005xt}, for slow-roll brane inflation rather than the DBI setup considered in this paper.\\

The structure of the paper is as follows.
In section \ref{sec:dbi} we summarise the salient details of DBI inflation relevant for our purposes, followed by a brief explanation
of BGMPZ backgrounds.
In section \ref{sec:calc} we explain the details of the numerical procedure adopted in solving the BGMPZ equations.
Example results are presented in section \ref{sec:results}, and in section \ref{sec:discuss} we discuss our results before concluding.

\section{Theoretical Background}
\subsection{DBI inflation}\label{sec:dbi}
The DBI inflation scenario takes place in a type IIB string theory whose ten--dimensional target space is compactified on a
Calabi--Yau threefold with at least one warped throat, containing a $\overline{\text{D3}}$ brane at its tip.
This then attracts a D3 brane from higher up in the throat, which rolls down the warped geometry.\\

An approximate ansatz for the throat geometry (which can be thought of as a warped AdS space) is $AdS_5\times T^{1,1}$,
with $T^{1,1}$ a five--dimensional Sasaki--Einstein manifold.
One may choose the metric to have the form\footnote{Here we work in the Einstein frame, which is best suited to the calculation
of cosmological observables. We will later switch to the string frame when discussing the BGMPZ solutions.}:
\begin{equation}
\label{eq:metric}
  ds^2=f^{-1/2}(r)dx^2+f^{1/2}(r)(dr^2+ds_5^2),
\end{equation}
where $dx^2$ describes our four--dimensional space--time; $r$ is a radial coordinate (i.e. the distance
from the tip of the throat); $f(r)$ the warp factor and $ds_5^2$ describes the Sasaki--Einstein manifold.
The moving D3 brane is described by its Dirac--Born--Infeld (DBI) action.
To write this down, we use the scaled radial coordinate $\phi=T^{1/2}r$ and warp factor $\tilde{f}(r)=T^{-1}f(r)$,
where $T$ is the string tension of the D3--brane. Then the action reads:
\begin{equation}
\label{eq:DBI}
  S=-\int dx^4\left(\tilde{f}^{-1}\left((1-\tilde{f}\dot{\phi}^2)^{1/2}-1\right)-V(\phi)\right).
\end{equation}
One now identifies $\phi$ with the inflaton field. This moves ultra--relativistically in general, but is subject to a speed limit $\dot{\phi}<\tilde{f}(\phi)^{-1}$ arising from reality of the action. The corresponding Lorentz factor for the D3 brane is:
\begin{equation}
\label{eq:gamma}
\gamma=\sqrt{1+4M_p^4H'^2\tilde{f}(\phi)},
\end{equation}
where $M_p$ is the Planck mass and $H(\phi)$ the Hubble parameter, with primes denoting differentiation with
respect to $\phi$.
The fact that the speed of the brane is limited means that the dominant contribution to its energy comes from
the potential $V(\phi)$, and it is this that allows one to identify the field $\phi$ with the inflaton.
One assumes the potential to be dominated by the quadratic mass term\footnote{Note that in recent explicit
constructions of D3--brane inflation with embedded D7--branes such a quadratic term does not appear in the
effective potential~\cite{Krause:2007jk,Baumann:2007ah}.},
and thus (following~\cite{Shandera:2006ax}) the Hubble parameter may be written:
\begin{equation}
\label{eq:h}
H(\phi)=m\phi(1-B\phi^2)/M_p,
\end{equation}
where $B$ represents the first in a series of small correction terms arising from the kinetic energy\footnote{This form of
the potential always fulfills the consistency relation of~\cite{Spalinski:2007dv}.}.
It is customary to introduce the following inflationary parameters:
\begin{equation}
\epsilon_D\equiv\frac{2M_p^2}{\gamma}\left(\frac{H'(\phi)}{H(\phi)}\right)^2,\quad \eta_D\equiv\frac{2M_p^2}{\gamma}\left(\frac{H''(\phi)}{H(\phi)}\right),\quad \kappa_D\equiv\frac{2M_p^2}{\gamma}\left(\frac{H'(\phi)}{H(\phi)}\frac{\gamma'(\phi)}{\gamma(\phi)}\right).\\
\label{slowroll}
\end{equation}
The scalar spectral index is given to linear order in these parameters by \cite{Shandera:2006ax}:
\begin{equation}
n_s-1\simeq -4\epsilon_D+2\eta_D-2\kappa_D.
\label{ns}
\end{equation}
Note that those parameters involving $H''(\phi)$ vanish for the Hubble parameter given above with $B=0$, but not in the more general case where higher order kinetic and potential energy corrections are included. The number of $e$--folds before the end of inflation is given by:
\begin{equation}
\label{eq:Ne}
N_e=\int H dt=-\frac{1}{2M_p^2}\int\frac{H}{H'}\gamma(\phi)d\phi.
\end{equation}
where the integral proceeds from initial to final values of the inflaton field $\phi$. Another quantity of interest is the ratio of the amplitudes of tensor and scalar fluctuations, which to linear order in the above parameters is given by~\cite{Shandera:2006ax}:
\begin{equation}
r\simeq \frac{16\epsilon_D}{\gamma}.
\label{r}
\end{equation}
\subsection{BGMPZ solutions}\label{sec:bgmpz}
The BGMPZ~\cite{Butti:2004pk} geometries arise as solutions in the supergravity limit of type IIB string theory, in the presence
of fluxes.
The explicit form of the solutions for this particular one--parameter family has been given in~\cite{Butti:2004pk} in the form of solutions
to a series of coupled first order ordinary differential equations describing the functions entering the metric tensor.
The KS and MN solutions were shown to emerge as particular cases of the BGMPZ solution, which is characterised by a 
parameter~$\xi$ which smoothly interpolates between these two extremal cases.
A previously found deformation of the KS geometry~\cite{Papadopoulos:2000gj} was also shown to emerge as a particular solution.\\

More specifically, the BGMPZ metric in the string frame is:
\begin{equation}
ds^2=e^{2A(t)}dx_\mu\,dx^{\mu}+e^{-6p(t)+x(t)}dt^2+ds_5^2,
\label{metric}
\end{equation}
with the first two factors belonging to $AdS_5$ and $ds_5^2$ describing the Sasaki--Einstein space (see~\eqref{eq:metric} for
the analog of this equation in the Einstein frame).
It is convenient to introduce the quantity:
\begin{equation}
v(t)=e^{6p(t)+2x(t)}
\label{v}
\end{equation}
and this together with the functions $A(t)$, $x(t)$ and the dilaton $\Phi(t)$ are all that is needed to examine the inflationary consequences of this geometry. These functions obey a set of non--linear coupled first--order differential equations with additional functions $h_2(t)$, $a(t)$ arising from the fluxes and metric in the compact internal space. A full analytic solution of this system of equations does not seem to be possible, but results including appropriate integration constants are obtainable as a power series at asymptotically small or large $t$. In particular, $a(t)$ has the form \cite{Butti:2004pk}:
\begin{align}
a(t)&\rightarrow -1+\xi t^2+{\cal O}(t^4),\quad t\rightarrow 0;\label{alowt}\\
&\rightarrow -2e^{-t}+a_{UV}(-1+t)e^{-\frac{5t}{3}}+{\cal O}(e^{-{\frac{7t}{3}}}),\quad t\rightarrow\infty.
\label{ahight}
\end{align}

This defines the parameter $\xi$ characterising the family of BGMPZ geometries. It lies in the range:
\begin{equation}
\frac{1}{6}\leq\xi\leq\frac{1}{2},
\label{xirange}
\end{equation}
with $\xi=1/6,$ $1/2$ corresponding to the MN and KS solutions respectively. The UV ($t\rightarrow\infty$) behaviour of the dilaton changes sharply as one approaches the MN solution. For $\xi\neq1/6$, it approaches a constant, instead diverging for the MN solution itself. The parameter $a_{UV}$ in equation (\ref{ahight}) is fixed by requiring a regular solution for the metric. Similar power series are given in \cite{Butti:2004pk} for the functions $A(t)$, $\Phi(t)$, $v(t)$, $h_2(t)$ and $x(t)$. The metric in the Einstein frame, required to make contact with cosmology, is given by:
\begin{equation}
ds_E^2=e^{\Phi(t)/2}ds^2,
\label{Einstein}
\end{equation}
where $\Phi(t)$, as stated above, is the dilaton field and not to be confused with the inflaton field $\phi$ introduced in previous sections. Combining equations (\ref{eq:metric}, \ref{metric}, \ref{Einstein}) one finds:
\begin{equation}
f(r)=e^{-4A(t(r))-\Phi(t(r))/2},
\label{fr}
\end{equation}
and:
\begin{equation}
r(t)=\int_0^t dt'\left[\frac{e^{2A(t')+x(t')+\Phi(t')/2}}{v(t')}\right]^{\frac{1}{2}}.
\label{rt}
\end{equation}
The integration limits for $r(t)$ are chosen to fulfil the boundary condition $r(0)=0$. From equations (\ref{eq:metric}, \ref{fr}, \ref{rt}) one sees that cosmological implications of the BGMPZ solution occur in two ways. Firstly, through the warp factor (including the dilaton from the frame transformation). Secondly, through the radial coordinate transformation $r(t)$. The procedure for obtaining cosmological observables is as follows. One first solves the BGMPZ equations in the string frame for the quantities introduced above. As discussed, this is not possible analytically and so a numerical solution must be performed. One then transforms the metric to the Einstein frame in order to calculate physical quantities. The details of the numerical solution are discussed in the following section.
\section{Numerical Solution of the BGMPZ Equations}
\label{sec:calc}
\subsection{String Frame Metric}
The BGMPZ equations are a set of coupled non--linear first--order differential equations in the string frame radial coordinate $t$, which are subject to a number of potential numerical instabilities around $t=0$. Here we outline our procedure for numerical solution, before discussing the transformation of the resulting metric to the Einstein frame.\\

\begin{figure}
\begin{center}
\includegraphics[width=0.75\linewidth]{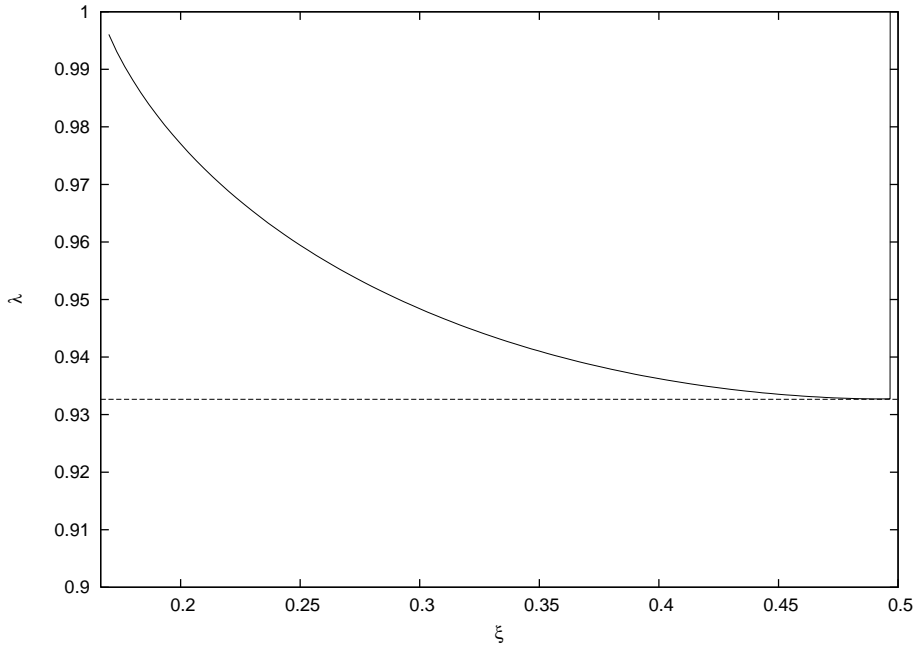}
\caption{Behaviour of the parameter $\lambda(\xi)$ defined in equation (\ref{lambda}), together with its asymptotic value $\lambda(1/2)=0.93266$. One sees that this value is reached before the numerical instability sets in.}
\label{lambdaplot}
\end{center}
\end{figure}

We solve the BGMPZ equations~\cite{Butti:2004pk} using a third order Adams--Bashforth algorithm. This is a multi--step method requiring 3 previous function evaluations at each integration step. Initial values are provided with a fourth order Runge--Kutta algorithm. A problem arises in that the analytic expressions for the derivatives of some of the metric functions are numerically unstable at $t=0$. This is easily solved by using the power series results for $a(t)$, $v(t)$ etc. in this regime, and matching at some suitable value $t=t_0$ chosen as small as possible so that the power series and full expressions for the derivatives are approximately equal. Furthermore, the power series for $A(t)$, $h_2(t)$ and $x(t)$ depend upon a parameter $\lambda(\xi)$ which depends on the values of the dilaton at $t=0$, $\infty$ ($\Phi_0$ and $\Phi_{UV}$ respectively)\footnote{One is free to choose the initial value of the dilaton, as in the supergravity approximation this amounts merely to an overall scaling of the metric. We set $\Phi_0=0$ for our analysis.}:
\begin{equation}
\lambda=\frac{2(1-e^{-2(\Phi_{UV}-\Phi_0)})^{\frac{1}{2}}}{3(1-2\xi)}.
\label{lambda}
\end{equation}
This appears to diverge at $\xi=1/2$. However, one then has the KS solution in which the dilaton is constant so that $\Phi_{UV}=\Phi_0$. Thus $\lambda(1/2)$ is in principle finite and is known to have a value $\lambda(1/2)=0.93266$~\cite{Butti:2004pk}. In practise $\Phi_{UV}$ must be found by numerical solution of the BGMPZ equations, and thus equation (\ref{lambda}) becomes unstable. We solve this by interpolating the numerically found $\lambda$ over a small region close to $\xi=1/2$ so that the asymptotic KS value is indeed reached. Numerical results for $\lambda(\xi)$ are shown in figure \ref{lambdaplot} with no such interpolation, and one sees that the asymptotic value is reached before the instability occurs, so that results near the KS solution will not be sensitive to the interpolation.
For large values of $t$ the numerical solutions for $A(t)$ and $x(t)$ also become unstable, and must be matched to the UV power series given in ~\cite{Butti:2004pk}. The series for $\exp(2A(t))$ has an overall $\xi$ dependent normalisation which must be found numerically from the region over which the numerical and power series solutions overlap. In addition, some interpolation is necessary between the numerical and power series solutions. If $A_n$ is the numerical solution and $A_p$ the power series, a suitable choice is:
\begin{align}
e^{2A}&\simeq (1-0.5e^{\alpha(t-t_1)})e^{2A_n}+0.5e^{\alpha(t-t_1)}e^{2A_p},\quad t<t_1;\notag\\
&\simeq 0.5e^{-\alpha(t-t_1)}e^{2A_n}+(1-0.5e^{-\alpha(t-t_1)})e^{2A_p},\quad t>t_1,
\label{inter}
\end{align}
with $\alpha$ selected to ensure a smooth interpolation. We have checked that final results for those quantities depending on the derivative of the warp factor are not sensitive to this interpolation. A further check on our numerical solutions can be performed by comparing to the known KS and MN solutions at $\xi=1/2$ and $\xi=1/6$ respectively. We have explicitly verified that the shape and normalisation of the analytical results for the above functions in these limits are reproduced.
\subsection{Transformation to the Einstein Frame}
Once the BGMPZ metric has been found, equations (\ref{fr}, \ref{rt}) provide numerical solutions for the Einstein frame warp factor and radial coordinate. The numerical solution for these functions consists of a series of values at discrete values of $t$. Values at general values of $t$ are obtained by interpolating between those on the discretised axis of $t$ points, and a linear interpolation is found to be sufficient. Equation (\ref{rt}) must be integrated numerically, and we use a 96--point Gaussian integration.\\

When calculating cosmological observables, one must find the value of the string frame coordinate $t$ that corresponds to a given value of $\phi\propto r(t)$. Thus the solution for $r(t)$ must be inverted numerically to find $t(r)$. We use a bisection algorithm with sufficiently many iterations. \\

Each of the numerical procedures described above introduces various tuning parameters (e.g. step sizes, interpolation parameters, iteration numbers) which control the convergence and accuracy of the final numerical results. We have checked the (in)sensitivity of all final results to these parameters.
\section{Results}
\label{sec:results}
\subsection{Warp Factors}
Several parameters must be fixed in order to produce numerical predictions from the above analysis. Firstly, the values of the string scale and inflaton mass. Secondly, there is an overall normalisation of the warp factor $\tilde{f}(\phi)$, which as well as the string scale and coupling has a dependence on the D3 brane tension and the deformation parameter of the conifold solution. Furthermore, there is a scaling factor ${\cal R}$ in the definition of the inflaton $\phi$, which also depends on the conifold deformation, brane tension and, in the case of inflation with higher dimensional wrapped branes~\cite{Kobayashi:2007hm,Becker:2007ui}, upon a scale factor involving the volume of the wrapped cycle.\\

Here we present results for the following choice of parameters:
\begin{equation}
m_s=0.01M_p;\quad m=10^{-6}M_p;\quad \tilde{f}(0)=10^7, \quad {\cal R}=25,
\label{paramchoice}
\end{equation}
where $m_s$ is the string scale (we work in units such that $m_s=1$), $m$ the inflaton mass and $\phi(t)={\cal R}r(t)$ . Our objective here is merely to demonstrate an example of how the numerical solution of the BGMPZ equations can be applied to calculate cosmological observables. We discuss in more detail the possible constraints on DBI models in section~\ref{sec:constraints}.\\

\begin{figure}
\begin{center}
\includegraphics[width=0.75\linewidth]{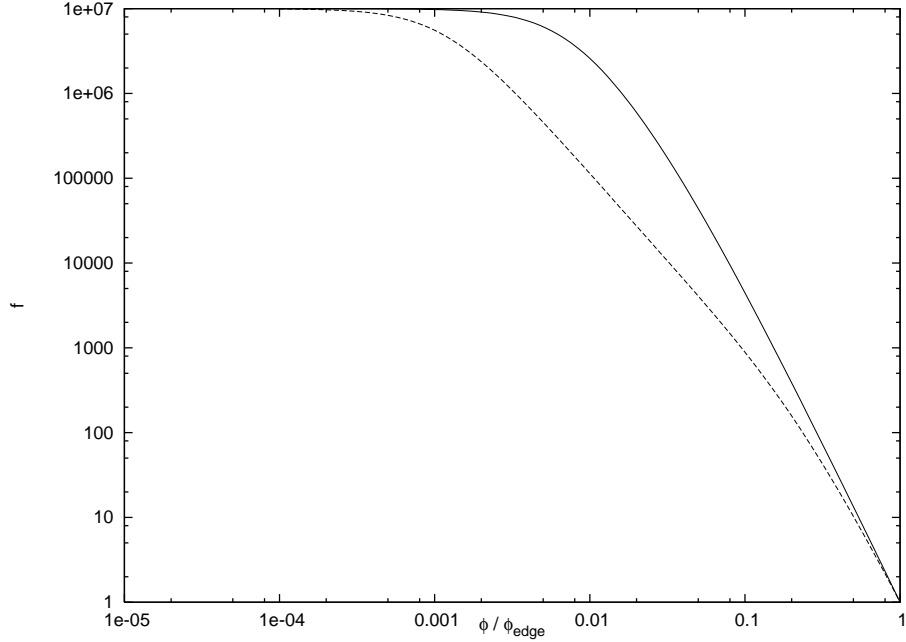}
\caption{Warp factors from the numerical solutions of the BGMPZ equations. Results are shown for $\xi=1/2$ (solid line) and $\xi=0.167$ (dashed line).}
\label{fplot}
\end{center}
\end{figure}

\begin{figure}
\begin{center}
\includegraphics[width=0.75\linewidth] {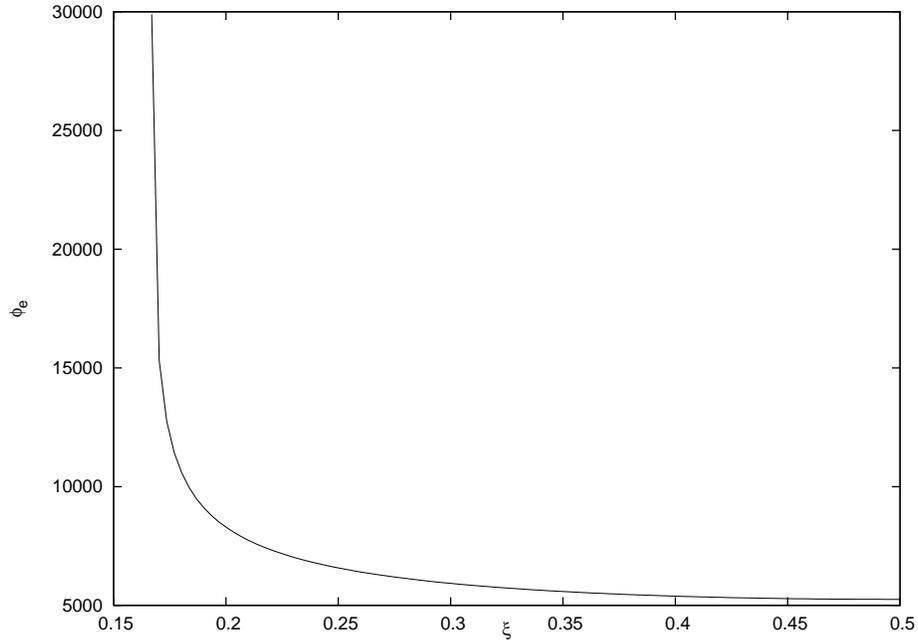}
\caption{Value of the inflaton $\phi_e$ corresponding to the radial coordinate at which the throat is glued to the bulk geometry, defined by $\tilde{f}(\phi_e)=1$.}
\label{phiexi}
\end{center}
\end{figure}

The BGMPZ warp factors for $\xi=1/2,$ $0.167$ (corresponding to the KS solutions and a solution close to the MN case respectively) are shown in figure \ref{fplot} as a function of $\phi/\phi_e$, where $\phi_e$ is the value of the inflaton field at which the warped throat joins onto the bulk geometry defined by:
\begin{equation}
f(\phi_{e})=1.
\label{phie}
\end{equation}
Thus, $\phi_e$ has a different value for each of the geometries. One sees this from figure~\ref{phiexi}, which shows $\phi_e$ as a function of $\xi$.
One sees that $\phi_e$ diverges towards the MN solution ($\xi\rightarrow 1/6$) as expected from the divergence of the dilaton $\Phi(t)$. Two things are evident from figure \ref{fplot}. Firstly, both warp factors converge for low values of $\phi$, corresponding to low values of the string frame radial coordinate $t$ where the geometries are the same. Constancy of the warp factor as $\phi\rightarrow 0$ demonstrates that the BGMPZ solutions do indeed have a dynamically generated cutoff of the warped throat. Secondly, the warp factors converge as $\phi/\phi_e\rightarrow 1$. This is a consequence of both geometries describing warped throats that are asymptotically AdS. \\

The difference between the warp factors in figure \ref{fplot}, which effectively represent extremal cases of the BGMPZ geometry, emerges in the shape at intermediate $\phi$ values. The KS throat ($\xi=1/2$) interpolates smoothly between the flat warp factor at low values of $\phi$, and the asymptotic AdS regime towards the tip. It can be effectively parameterised by a function of form~\cite{Kecskemeti:2006cg}:
\begin{equation}
f(\hat{\phi})=\frac{{\cal N}}{(\mu^2+\hat{\phi}^2)^2},\quad \hat{\phi}=\frac{\phi}{\phi_e},
\label{fpar1}
\end{equation}
for suitable parameters ${\cal N}$ and $\mu$. An extra qualitative regime is present in the general BGMPZ warp factor, namely a shoulder at intermediate $\phi$ values whose slope differs from the eventual asymptotic AdS behaviour. Thus, an extra parameter is needed in the denominator of equation (\ref{fpar1}), and an appropriate generalisation of the function (\ref{fpar1}) was also first suggested in~\cite{Kecskemeti:2006cg}:
\begin{equation}
f(\hat{\phi})=\frac{{\cal N}}{f_0+f_2\hat{\phi}^2+f_4\hat{\phi}^4}.
\label{fpar2}
\end{equation}
We have checked explicitly that this function is able to reproduce well the shape of the BGMPZ warp factor. Thus, the results of this paper confirm that phenomenological warp factors of form (\ref{fpar2}) do indeed correspond to geometries exactly realisable in string theory - namely the BGMPZ solutions.\\

\begin{figure}
\begin{center}
\includegraphics[width=0.75\linewidth,trim=25mm 30mm 30mm 15mm,clip]{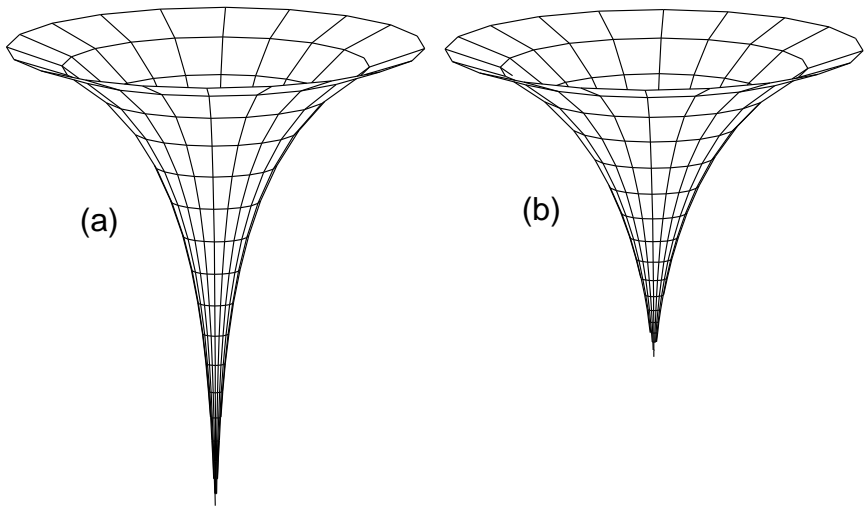}
\caption{Visualisation of the geometry of two throats at (a) $\xi=1/2$ and (b) $\xi=0.167$. The horizontal distance is in units of $\phi/\phi_e$, and the proper distance between two points at different values of the inflaton field is represented by the length of a geodesic path connecting them along the throat surface.}
\label{fig:throats}
\end{center}
\end{figure}

To give an impression of how one might visualise the two extremal cases at $\xi=1/2$ (KS) and $\xi=0.167$ (close to the MN solution), Figure~\ref{fig:throats}
shows two--dimensional slices of the geometry, parametrised by the radial parameter $\phi/\phi_e$. The vertical direction is used to show the warping, in such a way that the proper distance between two points at different radial values (given by equation (\ref{eq:metric})) is represented by the length of a geodesic path connecting them on the throat surface. The heights of the throats are scaled by $\phi_e$ for each case. The KS throat is then longer on such a plot than the MN--type solution, consistent with figure~\ref{fplot}, due to the higher warp factor. \\

\begin{figure}
\begin{center}
\includegraphics[width=0.75\linewidth]{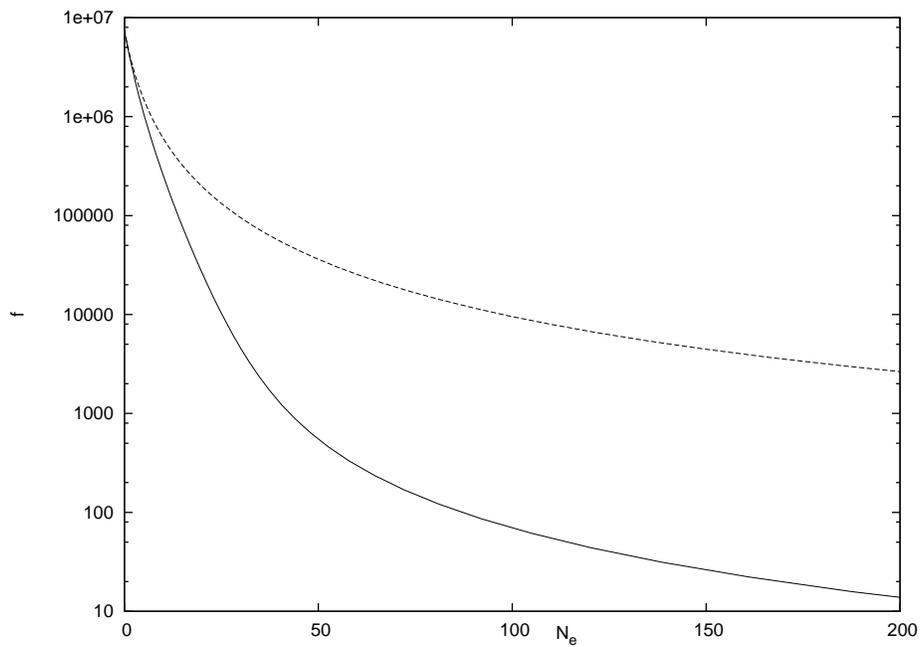}
\caption{Warp factors from the numerical solutions of the BGMPZ equations considered as a function of the number of $e$--folds before inflation ends, $N_e$. Results are shown for $\xi=1/2$ (solid line) and $\xi=0.167$ (dashed line).}
\label{fneplot}
\end{center}
\end{figure}

Although viewing the warp factor as a function of $\phi/\phi_e$ (as in figures~\ref{fplot} and~\ref{fig:throats}) allows one to more easily visualise the geometries, the physical scale on which to consider the warp factor is the number of $e$--folds. Thus, in figure~\ref{fneplot} we show the warp factors for $\xi=1/2$, $0.167$ as a function of $N_e$.
Contrary to the na\"{i}ve expectation of figure~\ref{fplot}, the warp factor at a given value of $N_e$ increases as one moves away from the KS solution. This is due to an increase in $\phi_e$ (and, thus, the value of $\phi$ at which a given $N_e$ occurs) as $\xi$ decreases. 

\subsection{Scalar Spectral Index}
\begin{figure}
\begin{center}
\includegraphics[width=0.75\linewidth]{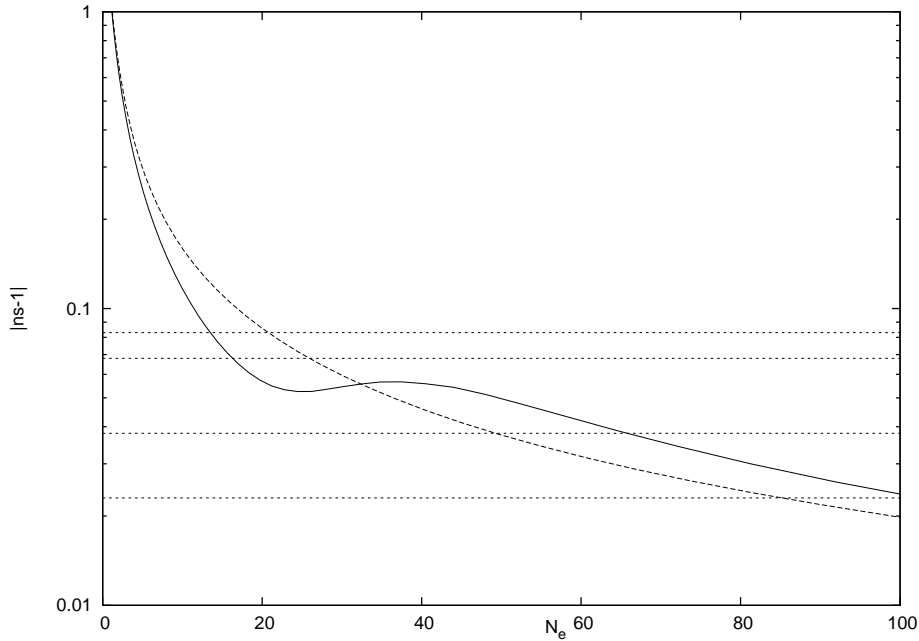}
\caption{Scalar spectral indices from the numerical solutions of the BGMPZ equations, followed by transformation to the Einstein frame. Results are shown for $\xi=1/2$ (solid line) and $\xi=0.167$ (dashed line). Also shown are 1$\sigma$ and $2\sigma$ error bands from the WMAP3 data.}
\label{nsneplot}
\end{center}
\end{figure}

In figure \ref{nsneplot} we show the spectral index as a function of the number of $e$--folds before inflation ends, defined by equation (\ref{eq:Ne}).  
Noticeably, the qualitative behaviour of the spectral index at high $e$--fold numbers differs between the two cases. The KS solution shows a bump which is absent as $\xi$ tends towards the MN background. Furthermore, this bump occurs (for our choice of parameters) within the last 50 $e$-folds of inflation. Superimposed on the plot are $1\sigma$ and $2\sigma$ error bands for the spectral index $n_s$ at 0.002/Mpc as measured from current data including WMAP3~\cite{Spergel:2006hy}:
\begin{equation}
n_s=0.947\pm 0.015.
\label{nsval}
\end{equation}
Note that this value is derived from a fit to combined data sets performed under the assumption that the ratio of tensor to scalar fluctuations, $r$, is zero. More properly one should should consider confidence contours in the ($r$, $n_s$) plane. Then, for non-zero $r$, the central value of $n_s$ increases but the uncertainty band has roughly the same width for reasonable variation of $n_s$ (see figure 14 in \cite{Spergel:2006hy}). In the DBI scenario, $r$ is indeed non-zero and varies with $\xi$. Nevertheless, we plot the value of equation (\ref{nsval}) in figure \ref{nsneplot} so as to indicate the amount of uncertainty encountered in current measurements of $n_s$. The uncertainty band in any case moves relative to the curves as a result of varying the DBI parameter space.\\

\begin{figure}
\begin{center}
\includegraphics[width=0.75\linewidth]{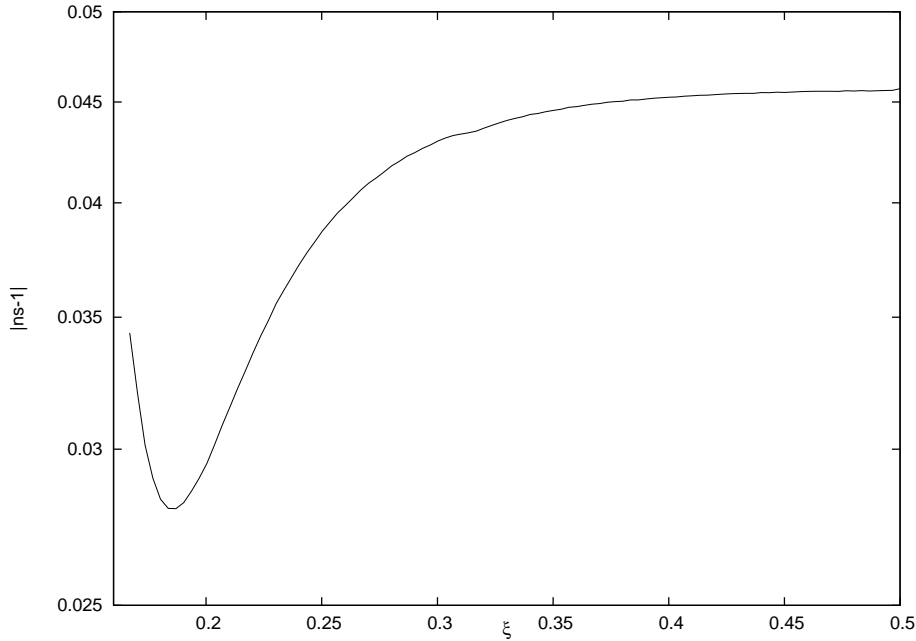}
\caption{Scalar spectral index 55 $e$--folds before the end of inflation, as a function of the BGMPZ parameter $\xi$.}
\label{nsxi}
\end{center}
\end{figure}

In figure \ref{nsxi} we show the value of $|n_s-1|$ at $N_e=55$, as a function of the $\xi$ parameter characterising the BGMPZ geometries.
For the $\xi=0.167$ case, one sees two ranges of $N_e$ in which the value of $|n_s-1|$ is less than and greater than the KS result respectively. As $\xi$ increases from 0.167, the crossing point of the two curves also varies such that for low $\xi$ it lies at $N_e<55$ whereas for higher $\xi$ it lies at $N_e>55$. Thus, for higher $\xi$ the value of $|n_s-1|$ increases with increasing $\xi$, whereas the opposite is true at low $\xi$. For points in the parameter space other than those of equation (\ref{paramchoice}), the spectral index may either increase or decrease as one moves away from the KS solution.\\

It is also interesting to note from figure~\ref{nsxi} that the value of $|n_s-1|$ at $N_e=55$ for the MN-like solution lies outside the $1\sigma$ WMAP3 error band such that is slightly disfavoured relative to the KS solution. For other points in the parameter space, this difference may be enhanced and thus the ability of current data to constrain the $\xi$ parameter along with the other parameters is not in doubt\footnote{We found examples of other points in parameter space where the KS result for $|n_s-1|$ lies outside the $2\sigma$ error band from WMAP3 whereas the MN result does not, although these choices of parameters were also ruled out by non-Gaussianity constraints, discussed in section~\ref{sec:constraints}.}. \\

Also of interest is the running of the spectral index. However, for the parameter space point chosen for our illustrative results, the running is sufficiently small for each geometry as to be unobservable. A plot of $dn_s / d\ln{k}$ for the KS solution can be seen, albeit for a different region of parameter space, in~\cite{Shiu:2006kj}.
\subsection{Ratio of tensor to scalar fluctuations}
\begin{figure}
\begin{center}
\includegraphics[width=0.75\linewidth]{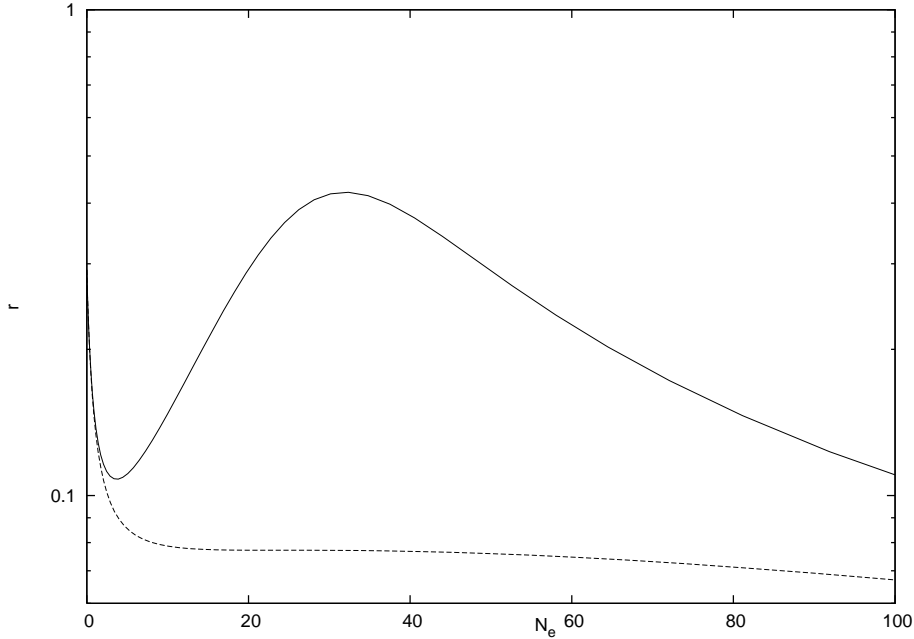}
\caption{Ratio of tensor to scalar modes from the numerical solutions of the BGMPZ equations, followed by transformation to the Einstein frame. Results are shown for $\xi=1/2$ (solid line) and $\xi=0.167$ (dashed line).}
\label{rneplot}
\end{center}
\end{figure}

\begin{figure}
\begin{center}
\includegraphics[width=0.75\linewidth]{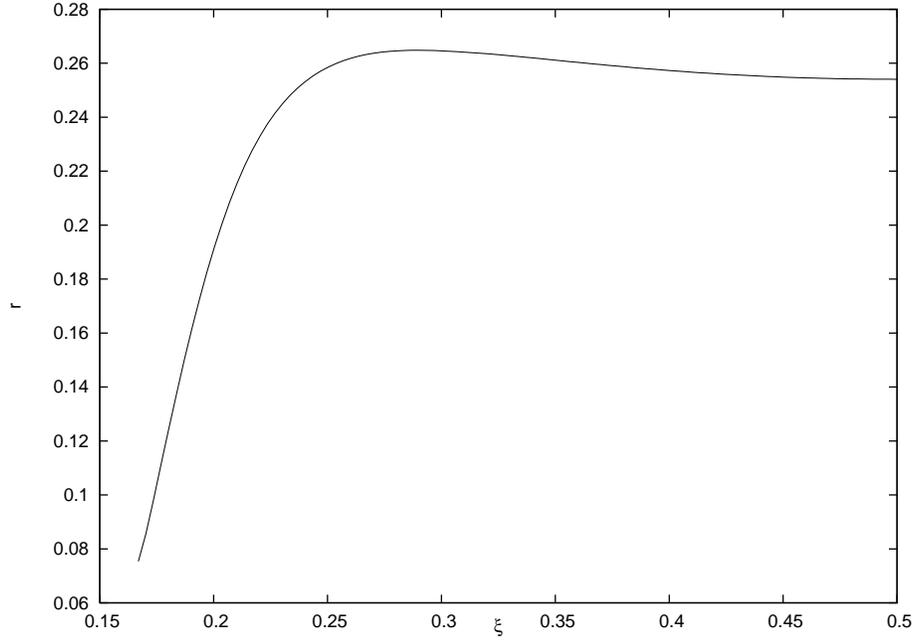}
\caption{Ratio of tensor to scalar modes 55 $e$--folds before the end of inflation, as a function of the BGMPZ parameter $\xi$.}
\label{rxi}
\end{center}
\end{figure}

One may also solve for the ratio of tensor to scalar modes $r$, given in equation (\ref{r}). Its behaviour as a function of the number of $e$--folds before the end of inflation is shown in figure \ref{rneplot} for $\xi=1/2$, $0.167$ where analogously to the case of the spectral index one sees a markedly different qualitative behaviour. The value of $r$ at $N_e=55$ is shown in figure \ref{rxi} as a function of $\xi$.
Similarly to the spectral index, one sees that $r$ may increase or decrease as $\xi$ increases.\\

\begin{figure}
\begin{center}
\includegraphics[width=0.75\linewidth]{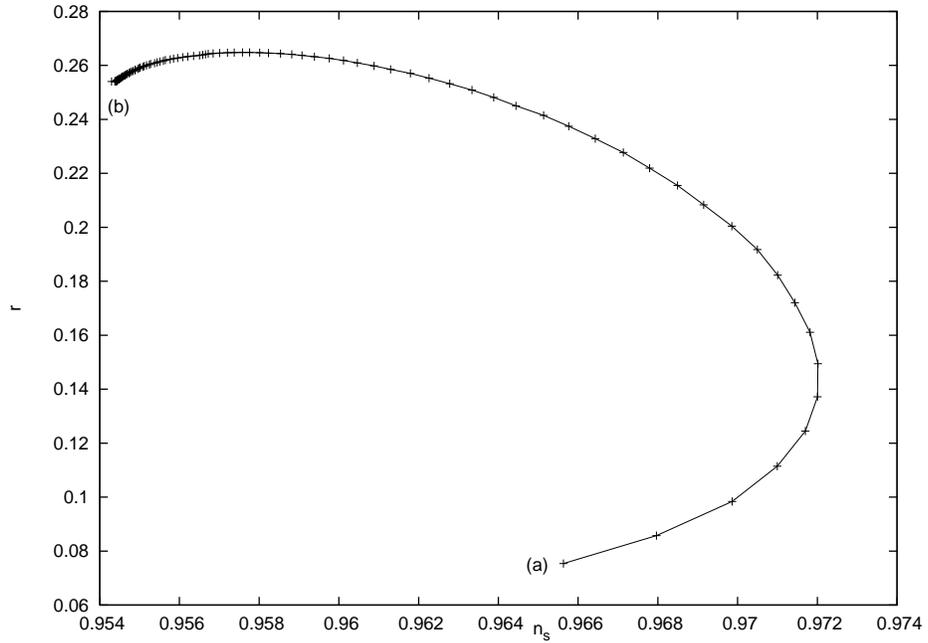}
\caption{Ratio of tensor to scalar modes vs. the scalar spectral index at 55 $e$-folds before inflation for (a) $\xi=0.167$; (b) $\xi=1/2$. Points on the curve are spaced equally in $\xi$.}
\label{rnsxiplot}
\end{center}
\end{figure}

Here we only wish to illustrate the sort of variation that can arise from the variation of the BGMPZ parameter, and the values of $r$ shown in figures \ref{rneplot} and \ref{rxi} cannot represent predictions without first ensuring that all constraints on DBI inflation are obeyed (see the following section for a discussion, and \cite{Baumann:2006cd} for a discussion of gravitational waves in DBI inflation). A full imposition of all theoretical and experimental constraints is beyond the scope of this paper. Nevertheless, in figure \ref{rnsxiplot} we show the value of $r$ versus $n_s$ at 55 $e$-folds before inflation, for various values of $\xi$ and using the parameters of equation (\ref{paramchoice}). Comparing with the corresponding figure in \cite{Spergel:2006hy}, one again reaches the conclusion that experimental data can constrain the BGMPZ parameter taken alongside the rest of the DBI parameter space. Theoretical constraints would act to further reduce the available parameter space. 

\subsection{Constraints on Parameters in DBI Inflation}
\label{sec:constraints}
In this paper we have presented results for a particular point in the parameter space of the DBI inflation model, namely that
of equation~(\ref{paramchoice}). 
This is merely to show that it is possible to calculate cosmological observables for this class of throat geometries, and also
the possibility that geometries with different $\xi$ values could in principle be differentiated by current and future data.
A much more thorough analysis would involve a scan over the parameter space including the various theoretical constraints that
are known to affect DBI inflation models, as it has been done for a generic form of the warping in~\cite{Bean:2007hc}.\\

There are a number of theoretical constraints which act on the parameter space of a given DBI inflation model.
Firstly, there is the requirement that the supergravity approximation used in deriving the geometry is valid, which includes the
condition that one may safely ignore the back--reaction of the colliding D--branes on the throat.
There is also the requirement that the throat fit inside the bulk geometry.
More practically, one must ensure that sufficiently many $e$--folds of inflation are generated.\\

\begin{figure}
\begin{center}
\includegraphics[width=0.75\linewidth]{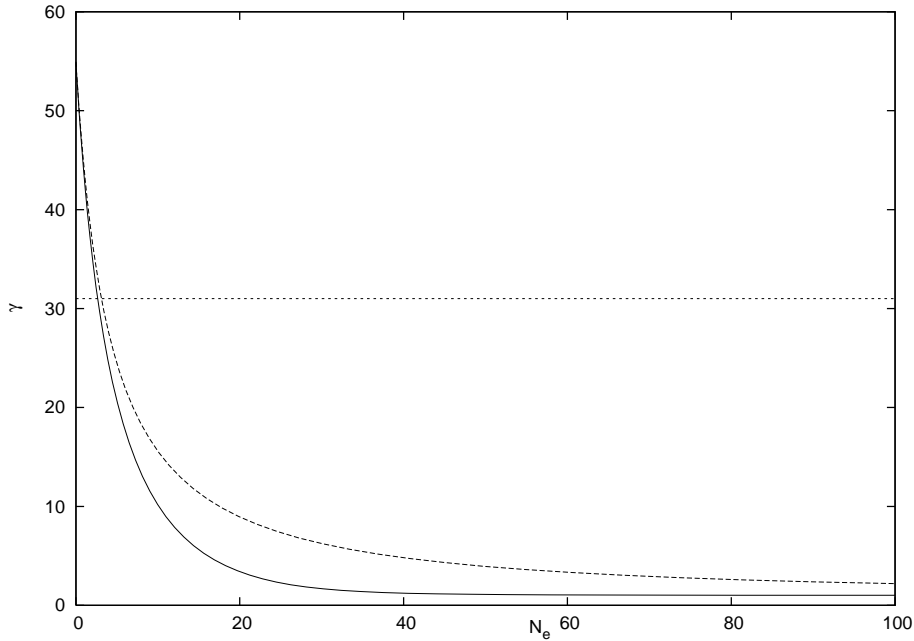}
\caption{Behaviour of the Lorentz factor $\gamma(\phi)$ as $N_e$ varies, for $\xi=1/2$ (solid) and $\xi=0.167$ (dashed). Also shown is the upper bound $\gamma(N_e\simeq 55)<31$ from non-Gaussianity constraints. }
\label{gammaplot}
\end{center}
\end{figure}

A further concern with DBI inflation in particular is that it generates a high degree of non--Gaussianity in the power spectrum of
cosmological perturbations, as a direct consequence of the large Lorentz factor associated with ultra--relativistic motion of the
moving D--brane~\cite{Alishahiha:2004eh,Chen:2006nt}.
It has been argued, for example, in~\cite{Baumann:2006cd,Bean:2007hc,Peiris:2007gz} that present non--Gaussianity measurements can be
used to heavily constrain the parameter space of the D3--$\overline{\text{D3}}$ model by requiring $\gamma(N_e=55)\lesssim 31$.
This is not a problem for the parameter choice of equation (\ref{paramchoice}), as seen in figure~\ref{gammaplot} where we show the behaviour of $\gamma(N_e)$ for $\xi=0.5$ and $\xi=0.167$.
One sees that both the KS and MN-like results lie well within the upper bound from non-Gaussianity, but that $\gamma(\phi)$ rises sharply as the number of $e$-folds decreases such that this constraint is a significant one. We also note that $\gamma(\phi)$ at a given $N_e$ increases as $\xi$ decreases, in other words that the non-Gaussianity increases as one moves away from the KS throat.\\

The non-Gaussianity constraint can be relaxed by considering generalisations of the original DBI scenario.
One such extension is to replace the D3 brane with higher dimensional D--branes that wrap cycles
in the throat~\cite{Kobayashi:2007hm,Becker:2007ui}, which might relax the previous constraints on the DBI inflation parameter space.
In principle, extended D--brane scenarios could be embedded in the family of throat geometries discussed in this paper, although the
question of how large an effect the back--reaction of higher dimensional branes has on the inflaton potential or in constraining the
DBI parameter space remains unclear\footnote{A preliminary study of back--reaction effects in higher dimensional brane inflation
scenarios is presented in \cite{Baumann:2007ah}.}.
Regarding non--Gaussianity, it is interesting that moving away from the KS throat results in a $\gamma$ factor which is higher for a given number of $e$--folds.
This suggests that analyses of DBI inflation based on the Klebanov--Strassler throat are underestimating possible non--Gaussianities.\\

Another influence on the parameters comes from the possibility that one could deal with multi--field inflation.
In this work we have only taken one direction of the D brane as the scalar inflaton field, but this could be extended to include more
directions.
As discussed recently in~\cite{Easson:2007dh,Huang:2007hh}, it can transpire that multi--field models are well--approximated
by single--field limits, which can also have an effect in relaxing previously given constraints on DBI models. \\

One has to be careful however assuming that it is possible to vary the parameter $\xi$ completely independently.
A priori the potential will change as we deform the geometry and therefore the ranges of all other parameters will
be influenced. In our simplified approach, we did not take these effects into account.\\

In light of the uncertainties above, the example parameters (\ref{paramchoice}) used in this paper may or may not be ruled out by
present measurements.
However, we seek to demonstrate only the fact that cosmological observables do depend quantitatively on the deformation parameter
$\xi$ of the BGMPZ geometries, and it is still very likely to be the case that generalised DBI setups can be approximated to a
reasonable degree by a throat of this kind.\\

\section{Discussion}\label{sec:discuss}
In this paper we have explicitly demonstrated the possibility of calculating cosmological observables in the DBI inflation setup
using a one parameter family of type IIB supergravity solutions that describe the geometry of a warped throat, the BGMPZ solutions
of Butti et al.~\cite{Butti:2004pk} that interpolate smoothly between the Klebanov--Strassler and the Maldacena--Nu{\~n}ez solution.
We have provided examples of cosmological parameters, namely spectral indices, that can be calculated from the underlying geometry.\\

The solution for the metric of the geometries in question is not possible analytically.
Therefore numerical methods have been used and shown to provide an adequate representation of the metric in terms of numerical
precision.
Instabilities in the derivatives entering the equations have been dealt with by matching to known power series expansions of the
metric functions at asymptotically small and large values of the radial coordinate.\\

We presented warp factors for two almost extremal ends of the family of solutions parametrised by~$\xi$:
the Klebanov--Strassler throat ($\xi=1/2$), and a geometry close to the Maldacena--Nu{\~ n}ez solution ($\xi=0.167$).
The qualitative behaviour of the warp factors is seen to be different.
Both solutions show a flat warp factor at low values of the rescaled radial coordinate $\phi$, corresponding to a dynamically
generated cutoff, and an asymptotically AdS behaviour at large $\phi$ values as expected.
However, the warp factor as it moves away in $\xi$ from the KS solution develops a shoulder at intermediate values, whose
slope is in general different from the eventual asymptotic behaviour.
This can be effectively parameterised by a function of form (\ref{fpar2}), thus explicitly demonstrating that phenomenological
fits of DBI inflation to cosmological data that assume a warp factor of this form indeed correspond to exactly realisable warped
throat geometries in a known string theory compactification.\\

We presented examples of scalar spectral indices and the ratio of tensor to scalar modes calculated using the different geometries.
The different qualitative behaviour observed in the warp factors carries through to the spectral indices, and a
quantitative difference between the values of the indices from different solutions at 55 $e$--folds before the
end of inflation is potentially measurable.

Constraints on non-Gaussianity and measured values of the scalar spectral index could already be used to rule out
some regions of $\xi$, when it is considered alongside the rest of the DBI parameter space.
Very generically, we have found that the amount of non--Gaussianity increases as one moves away from the KS solution (all other parameters being equal).\\

Since the warping changes qualitatively as one varies $\xi$, it is also conceivable that that certain constraints on the
parameter space may be relaxed by including $\xi$ as an additional parameter with respect to existing analyses such
as~\cite{Lorenz:2007ze}.
This issue certainly deserves further study.

\subsection*{Acknowledgements}
We would like to thank Gary Shiu for initiating this project and for very helpful correspondence.
We acknowledge informative discussions with Marieke Postma.
Our research is supported by the Dutch Foundation for Fundamental Research of Matter (FOM).

\bibliography{refs_inflation}
\bibliographystyle{utphys}

\end{document}